# Equation of state of single-crystal cubic boron phosphide

Y. Le Godec,[a,*] M. Mezouar,[b] O. O. Kurakevych,[c] P. Munsch,[c] U. Nwagwu,[d] J. H. Edgar,[d] and V. L. Solozhenko [e,**]

[a] *CNRS, UMR 7590, Paris, France*
[b] *European Synchrotron Radiation Facility, Grenoble, France*
[c] *IMPMC, Sorbonne Universités - UPMC Univ Paris 06, Paris, France*
[d] *Department of Chemical Engineering, Kansas State University, Manhattan KS, USA*
[e] *LSPM–CNRS, Université Paris Nord, Villetaneuse, France*

\* *e-mail: yann.legodec@impmc.upmc.fr*
\*\* *e-mail: vladimir.solozhenko@univ-paris13.fr*

**Abstract** – The 300 K equation of state of cubic (zinc-blende) boron phosphide BP has been studied by in situ single-crystal X-ray diffraction with synchrotron radiation up to 55 GPa. The measurements have been performed under quasi-hydrostatic conditions using a Ne pressure medium in a diamond anvil cell. A fit of the experimental *p-V* data to the Vinet equation of state yields the bulk modulus $B_0$ of 179(1) GPa with its pressure derivative of 3.3(1). These values are in a good agreement with previous elastic measurements, as well as with semiempirical estimations.

There are two boron phosphides: boron phosphide (BP) and icosahedral boron subphosphide ($B_{12}P_2$) [1, 2]. Both are refractory and wide bandgap semiconductors, and have attracted considerable interest for their superior physical properties, such as outstanding high-temperature stability and high thermoelectric power for direct energy conversion. These properties suggest many uses in modern technology. For example, they would be promising materials for large-surface-area liquid junction solar cells. Also, BP is characterized by a unique combination of properties that could make it a material of choice for a wide range of engineering applications [2]. BP shows promising mechanical [3], thermal, and electrical properties, excellent thermal conductivity (4 W cm$^{-1}$ deg$^{-1}$) [4] and a high thermoelectric power. The high thermal neutron capture cross-section of the $^{10}$B isotope makes boron phosphide appealing for a solid state neutron detector [5, 6]. Furthermore, the lattice parameter of BP closely matches that of zinc-blende gallium nitride, making it a suitable buffer layer on silicon for GaN light emitting diodes and laser diodes [7].

Although the promising aspect of extending injection luminescence to materials with large band gap has stimulated interest in BP, its high melting point combined with its high phosphorus vapor pressure makes growing of the single crystals difficult [8, 9]. The application of high pressure could, from one side, allow avoiding the oxidation and facilitate the growth of BP crystals, and, from other side, result in the synthesis of new high-pressure phase(s) in the B–P system. Thus, the equation of state as well as high-pressure thermal expansion data are of great importance for synthesis of this advanced material. However, at present the high-pressure behavior of boron phosphides is poorly studied and the available information is contradictory [10–12]. Here we report the equation of state data for single-crystal BP up to 55 GPa.

Single crystals of BP with sizes up to 100 µm were obtained using flux growth technique by interaction of phosphorus vapor with boron dissolved in nickel in a sealed (previously evacuated) quartz tube [13]. The boron-nickel solution was located at one end of the tube and held at 1150°C, while the phosphorus, which was initially located at the opposite end, was heated to 430°C (to produce a vapor pressure up to 0.5 MPa). Transparent red crystals were obtained with a cooling rate of 3 K/h. The lattice constant of the crystals was 4.534 Å, close to the literature data ($a = 4.538(2)$ Å [14]).

In situ X-ray diffraction experiments in a diamond anvil cell (DAC) were conducted at ID27 beamline of the European Synchrotron Radiation Facility (ESRF). A large-aperture membrane-type DAC with anvil tips 250 µm in diameter has been used. A small (~ 15 µm) single crystal of BP was loaded together with a small ruby ball (less than 5 µm in diameter) into the 90 µm diameter hole drilled in a rhenium gasket of 250 µm thickness preindented down to 28 µm. The BP crystal and pressure marker were placed within a few micrometers to each other close to the center of a diamond culet. Neon pressure medium has been used to maintain quasi-hydrostatic conditions. Pressure was determined in situ from the calibrated shift of the ruby $R_1$ fluorescent line [15] and equation of state of neon [16]. High-brilliance focused ($3 \times 3$ µm$^2$) synchrotron radiation was set to a wavelength of 0.3738(1) Å. High-pressure X-ray diffraction patterns were collected using on-line large-area Bruker CCD detector up to 55 GPa with exposure times varying from 3 to 5 min. The diffraction patterns were processed with FIT2D and General Structure Analysis System (GSAS) software.

Up to 40 GPa both pressure gauges (ruby, neon) indicated very close pressures, which points to the negligible strains and stresses, as well as inessential pressure gradients all over the cell. At higher pressures, Ne equation of state has been used as a pressure gauge. Single-crystal reflections (Fig. 1) were individually integrated after refinement of the beam center. The small differences between apparent lattice parameters for different Bragg lines (111 and 200 of Ne, 220 and 331 of BP) indicated the quasi-hydrostatic conditions during the measurements up to 55 GPa. Also, no evidence for any phase transformation to another crystalline structure of BP, or to an amorphous phase has been found over the pressure range studied.

Three theoretical equations of state were used to establish isothermal bulk modulus $B_0$ and its first pressure derivative $B_0'$, i.e. those of Vinet [17], Birch-Murnaghan [18], and Murnaghan [19]. The fitting results are listed in the table, while the Vinet fit is presented in Fig. 2. In the compression range probed here, all three models fit the data equally well and give almost the same values for $B_0$ and $B_0'$.

The fitted value of bulk modulus ($B_0 = 179(1)$ GPa with Vinet fit) well agrees with those obtained by elastic measurements ($B_0 = 173$ GPa) [20] and semi-empirical estimations ($B_0 = \sim 180$ GPa) [21, 22]; while it is somewhat higher (8.5 %) than the $B_0$ value estimated from the Cohen relationship between bulk moduli and interatomic A–B distances of diamond-like $A^XB^{8-X}$ compounds ($B_0 = 165.3$ GPa) [23]. In contrast, the previously reported experimental $B_0$ value of 267 GPa [24] (derived from a non-hydrostatic experiment without pressure medium) seems to be highly overestimated.

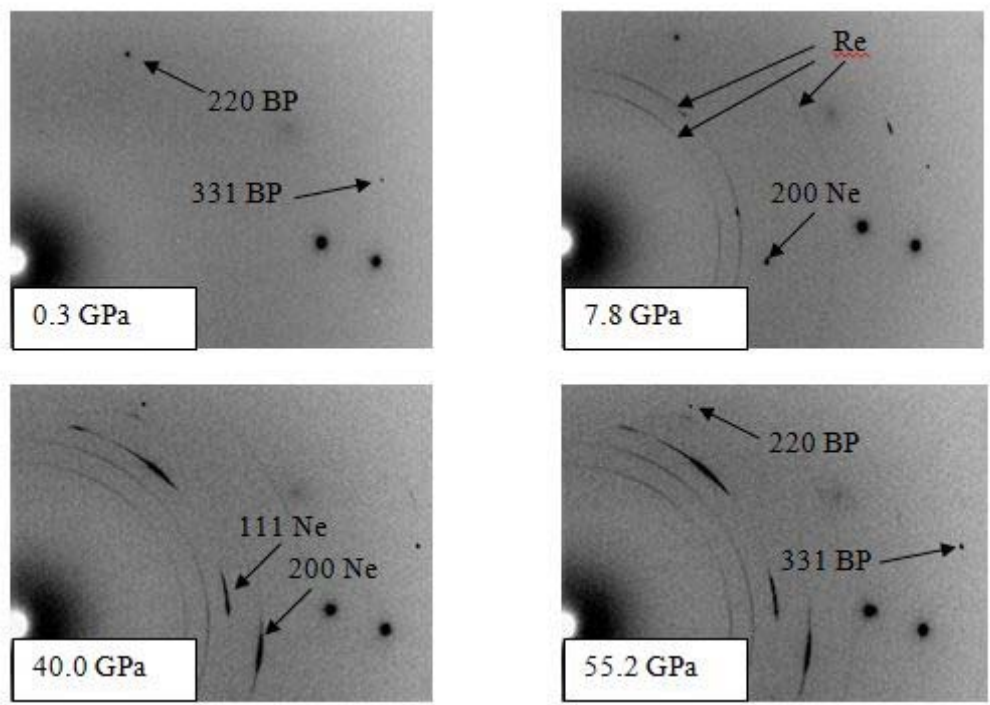

**Fig. 1**   X-ray diffraction patterns of single-crystal BP in Ne pressure medium at different pressures. The rhenium reflections arise from the gasket material.

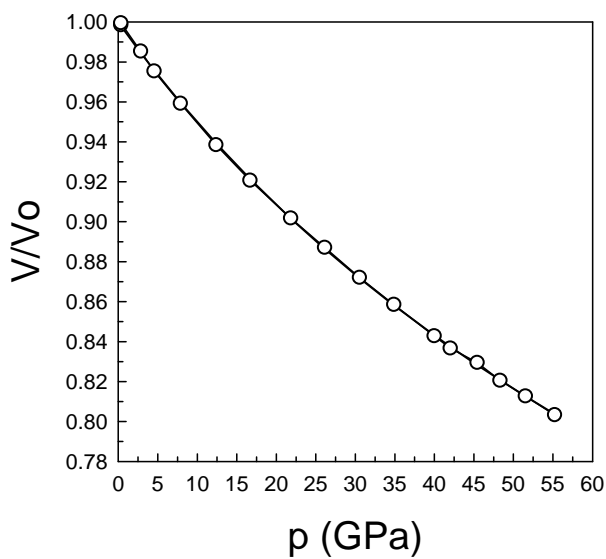

**Fig. 2**   300 K equation of state data for single-crystal BP (o – present work, solid line – fit to the Vinet EoS).

**Table**   Comparison of equation-of-state data of cubic BP to various EoS [17-19]. The zero-pressure volume $V_0$ was fixed to 93.2061Å$^3$

| Model | Vinet | Birch-Murnaghan | Murnaghan |
|---|---|---|---|
| BP (this study) | $B_0 = 179(1)$ GPa $B_0' = 3.3(1)$ | $B_0 = 178(1)$ GPa $B_0' = 3.3(1)$ | $B_0 = 180(1)$ GPa $B_0' = 3.0(1)$ |

The authors thank Dr. G. Garbarino for his help in preparing high-pressure experiments. This work was carried out at beamline ID27 during beamtime kindly provided by ESRF and was financially supported by the Agence Nationale de la Recherche (grant ANR-2011-BS08-018). The support for growing boron phosphide crystals from the U.S. Department of Energy (grant no. DE-SC0005156) is greatly appreciated.

REFERENCES

1. Amberger, E. and Rauh, P.A., Struktur des borreichen Borphosphide, *Acta Crystallogr. B,* 1974, vol. **30**, no. 11, pp. 2549–2553.

2. Popper, P. and Ingles, T.A., Boron Phosphide, a III–V Compound of Zinc-Blende Structure, *Nature*, 1957, vol. 179, no. 4569, pp. 1075–1075.

3. Mukhanov, V.A., Kurakevych, O.O., and Solozhenko, V.L., Thermodynamic Model of Hardness: Particular Case of Boron-Rich Solids, *J. Superhard Mater.*, 2010, vol. 32, no. 3, pp. 167–176.

4. Kumashiro, Y., Mitsuhashi, T., Okaya, S., Muta, F., Koshiro, T., Takahashi, Y., and Mirabayashi, M., Thermal Conductivity of a Boron Phosphide Single-Crystal Wafer up to High Temperature, *J. Appl. Phys.*, 1989, vol. 65, no. 5, pp. 2147–2148.

5. Lund, J.C., Olschner, F., Ahmed, F., and Shah, K.S., Boron Phosphide on Silicon for Radiation Detectors, *Mater. Res. Soc. Symp. Proc.*, 1989, vol. 162, no. 1, pp. 601–604.

6. Abbott, J.K.C., Brasfield, J.D., Rack, P.D., Duscher, G.J., and Feigerle, C.S. Chemical Vapor Deposition of Boron Phosphide Thin Films, *MRS Online Proceedings Library*, 2012, vol. 1432, no. 1, pp. 65–70.

7. Nishimura, S., Hirai, M., Nagayoshi, H., and Terashima, K., Strong Visible Light Emission from Zinc-Blende InGaN/GaN pn Junction on Silicon Substrate, *ECS Trans.*, 2013, vol. 53, no. 2, pp. 87–91.

8. Mukhanov, V.A., Sokolov, P.S., Le Godec, Y., and Solozhenko, V.L., Self-Propagating High-Temperature Synthesis of Boron Phosphide, *J. Superhard Mater.*, 2013, vol. 35, no. 6, pp. 415–417.

9. Kumashiro, Y., Refractory Semiconductor of Boron Phosphide, *J. Mater. Res.*, 1990, vol. 5, no. 12, pp. 2933–2947.

10. Badi, N., Amrane, N., Abid, H., Driz, M., Soudini, B., Khelifa, B., and Aourag, H., Pressure-Dependent Properties of Boron Phosphide, *Phys. Stat. Solidi B*, 1994, vol. 185, no. 2, pp. 379–388.

11. Amrane, N., Pressure Dependence of Positron Annihilation in Boron Phosphide, *Superlattices and Microstructures,* 2003, vol. 33, issues 1–2, pp. 9–21.

12. Zaoui, A., Kacimi, S., Yakoubi, A., Abbar, B., and Bouhafs, B., Optical Properties of BP, BAs and BSb Compounds under Hydrostatic Pressure, *Phys. B*, 2005. vol. 367, no. 1–4, pp. 195–204.

13. Nwagwu, U., Flux Growth and Characteristics of Cubic Boron Phosphide, *Master of Sci. Thesis*, Department of Chemical Engineering, Manhattan KS: Kansas State University, 2013, p. 87.

14. Rundqvist, S., Crystal Structure of Boron Phosphide BP, *Congres International de chimie pure et applique*, 16$^{eme}$ Paris 1957, Mem. Sect., 1958, vol. 1957, no. 1, pp. 539–540.

15. Mao, H.K., Xu, J. and Bell, P.M., Calibration of the Ruby Pressure Gauge to 800 kbar under Quasi-Hydrostatic Conditions, *J. Geophys. Res.*, 1986, vol. 91, no. B5, pp. 4673–4677.

16. Hemley, R.J., Zha, C.S., Jephcoat, A.P., Mao, H.K., Finger, L.W., and Cox, D.E., X-ray Diffraction and Equation of State of Solid Neon to 110 GPa, *Phys. Rev. B*, 1989. vol. 39, no. 16, art. 11820.

17. Vinet, P., Ferrante, J., Smith, J.R., Rose, J.H. A Universal Equation of State for Solids, *J. Phys. C*, 1986, vol. 19, no. 20, pp. L467–L473.

18. Birch, F., Finite Strain Isotherm and Velocities for Single-Crystal and Polycrystalline NaCl at High Pressures and 300-degree-K, *J. Geophys. Res.*, 1978, vol. 83, NB 3, pp. 1257–1268.

19. Murnaghan, F.D., The Compressibility of Media Under Extreme Pressures, *Proc. Natl. Acad. Sci.USA*, 1944, vol. 30, no. 9, pp. 244–247.

20. Wettling, W. and Windscheif, J., Elastic Constants and Refractive Index of Boron Phosphide, *Solid State Comm.*, 1984, vol. 50, no. 1, pp. 33–34.

21. Mukhanov, V.A., Kurakevych, O.O., and Solozhenko, V.L., The Interrelation between Hardness and Compressibility of Substances and Their Structure and Thermodynamic Properties, *J. Superhard Mater.*, 2008, vol. 30, no. 6, pp. 368–378.

22. Mukhanov, V.A., Kurakevych, O.O., and Solozhenko, V.L., Thermodynamic Aspects of Materials Hardness: Prediction of Novel Superhard High-Pressure Phases, *High Press. Res.*, 2008, vol. 28, no. 4, pp. 531–537.

23. Cohen, M.L., Calculation of Bulk Moduli of Diamond and Zinc-Blende Solids, *Phys. Rev. B*, 1985, vol. 32, no. 12, art. 7988.

24. Suzuki, T., Yagi, T., Akimoto, S.-i., Kawamura, T., Toyoda, S., and Endo, S., Compression Behavior of CdS and BP up to 68 GPa, *J. Appl. Phys.*, 1983, vol. 54, no. 2, pp. 748–751.